\documentclass[twocolumn,english,aps,prl,showpacs,superscriptaddress,groupaddress,floatfix,graphics,graphicx]{revtex4-1}
\usepackage[utf8]{inputenc}
\usepackage{color}
\usepackage{babel}
\usepackage{amsmath}
\usepackage{amssymb}
\usepackage{graphicx}
\usepackage{esint}
\usepackage[unicode=true,
 bookmarks=true,bookmarksnumbered=false,bookmarksopen=false,
 breaklinks=false,pdfborder={0 0 1},backref=false,colorlinks=false]
 {hyperref}
\hypersetup{
 colorlinks,linkcolor=blue,citecolor=blue,urlcolor=blue}

\makeatletter
\usepackage{babel}
\usepackage{babel}
\usepackage{babel}
\usepackage{babel}
\usepackage{babel}
\usepackage{babel}
\usepackage{mathrsfs}

\makeatother

\begin{document}

\title{Crystal-field effects in graphene with interface-induced spin\textendash orbit
coupling}
\begin{abstract}
We consider theoretically the influence of crystalline fields on the
electronic structure of graphene placed on a layered material with
reduced symmetry and large spin\textendash orbit coupling (SOC). We
use a perturbative procedure combined with the Slater-Koster method
to derive the low-energy effective Hamiltonian around the $K$ points
and estimate the magnitude of the effective couplings. Two simple
models for the envisaged graphene\textendash substrate hybrid bilayer
are considered, in which the relevant atomic orbitals hybridize with
either top or hollow sites of the graphene honeycomb lattice. In both
cases, the interlayer coupling to a crystal-field-split substrate
is found to generate highly anisotropic proximity spin\textendash orbit
interactions, including in-plane 'spin--valley' coupling. Interestingly, when an anisotropic intrinsic-type SOC becomes sizeable, the bilayer system is effectively
a quantum spin Hall insulator characterized by in-plane helical edge states robust against Bychkov-Rashba effect. Finally, we discuss the
type of substrate required to achieve anisotropic proximity-induced SOC and suggest possible candidates to further explore crystal field effects in graphene-based heterostructures. 
\end{abstract}

\author{Tarik P. Cysne}
\email{tarik.cysne@gmail.com}

\affiliation{Instituto de Física, Universidade Federal do Rio de Janeiro, Caixa
Postal 68528, Rio de Janeiro 21941-972, RJ, Brazil}

\author{Aires Ferreira}
\email{aires.ferreira@york.ac.uk}

\affiliation{Department of Physics, University of York, York YO10 5DD, United
Kingdom}

\author{Tatiana G. Rappoport}

\affiliation{Instituto de Física, Universidade Federal do Rio de Janeiro, Caixa
Postal 68528, Rio de Janeiro 21941-972, RJ, Brazil}
\maketitle

\section{Introduction}

The impact of the crystal environment on atomic states is pivotal
to understand the electronic structure \textcolor{black}{of solids}
containing transition metal atoms \cite{FazekasBook}. For instance,
in high-Tc cuprates, crystal field states are essential in the description
of $\text{CuO}_{2}$ planes, where $\text{Cu}^{+2}$ ions are surrounded
by elongated octahedral structures of $\text{O}$ atoms \cite{Dagotto1994,ZhangRice1988}.
Crystal electric field effect and its interplay with spin\textendash orbit
coupling plays an important role in magnetic anisotropy \cite{Sati2006,Sachidanandam97}
Jahn-Teller effect \cite{Jahn1936,Ishihara97,Salamon2001}, distortive
order \cite{Ishizuka1996} and cooperative Jahn-Teller effect \cite{Gehring1975}.

More recently, it has been appreciated that crystal field effect
(CFE) underlies rich spin-dependent phenomena at metallic interfaces.
For instance, the broken rotational symmetry of magnetic atoms in
metal bilayers was found to render spin currents anisotropic \cite{Cahaya2017},
while a staggered CFE associated to nonsymmorphic structures of metal
species is responsible for a giant enhancement of Rashba effect in
BaNiS$_{2}$ \cite{Cottin16}. Here, we investigate the electronic
properties of graphene placed on nonmagnetic substrates characterized
by a sizable CFE. Graphene\textendash substrate hybrid bilayers are
currently attracting enormous interest due to the combination of Dirac
fermions and prominent interfacial spin\textendash orbital effects
in the atomically-thin (two-dimensional) limit \cite{Song2010,Lee2015,Rajput16}.
Monolayers of group VI transition metal dichalcogenides (TMDs) are
a particularly suitable match to graphene as a high SOC substrate.
The peculiar spin\textendash valley coupling in the TMD electronic
structure \cite{Xiao_TMD,Wang_TMD_12,Mak_TMD_16} provides a unique all-optical
methods for injection of spin currents across graphene--TMD interfaces \cite{Muniz_Sipe_15,Fabian2015}, as recently demonstrated \cite{AvsarOzyilmaz2017,Luo_17}.\textcolor{black}{{}
Furthermore, the proximity coupling of graphene to a TMD base 
breaks the sublattice symmetry of pristine graphene leading to competing
spin\textendash valley and }Bychkov-Rashba \textcolor{black}{spin\textendash orbit
interactions} \cite{Avsar2014,WangMorpurgo2015,Whang2017,Zihlmann2018,Cummings2017,vanWees2017,velenzuelaNat2017,OmarVanWees2018}\textcolor{black}{.
The enhanced SOC paves the way to }\textcolor{black}{\emph{bona fide}}\textcolor{black}{{}
relativistic transport phenomena in systems of two-dimensional Dirac fermions,
including the inverse spin galvanic effect \cite{Milletari2017,Offidani2017}}.

On a qualitative level, the band structure of graphene weakly coupled
to a high SOC substrate can be understood from symmetry. The intrinsic
spin\textendash orbit coupling (SOC) of graphene is invariant under
the full symmetries of the point group $D_{6h}$, which includes 6-fold
rotations and mirror inversion about the plane \cite{KaneMele1}.
The reduction of the full point group in heterostructures is associated
with the emergence of other interactions \cite{Pachoud2014,Kochan_17}.
For example, interfacial breaking of inversion symmetry reduces the
point group $D_{6h}\rightarrow C_{6v}$, allowing finite (non-zero)
Bychkov-Rashba SOC \cite{Rashba2009}. The low-energy Hamiltonian
compatible with time-reversal symmetry is 
\begin{eqnarray}
\mathcal{H}_{C_{6v}} & = & \;\hbar\,v\left(\tau_{z}k_{x}\sigma_{x}+k_{y}\sigma_{y}\right)+\lambda_{{\rm {KM}}}\,\sigma_{z}\tau_{z}s_{z}\nonumber \\
 &  & +\,\lambda_{\text{R}}\,\left(s_{x}\sigma_{y}-\tau_{z}s_{y}\sigma_{x}\right)\,,\label{Hc6v}
\end{eqnarray}
where $v$ is the Fermi velocity of massless Dirac fermions, $\mathbf{k}=(k_{x},k_{y})$
is the wavevector around a Dirac point (valley), and $\tau_{i},\sigma_{i}$
and $s_{i}$ are Pauli matrices acting on valley, sublattice, and
spin spaces, respectively. Here, $\lambda_{{\rm {KM}}}$ ($\lambda_{{\rm {R}}}$)
are the energy scales of the intrinsic-type SOC (Bychkov-Rashba) interaction
enhanced by the proximity effect.

In addition, the interaction of graphene with an atomically flat substrate
renders the two carbon sublattices inequivalent, further reducing
the point group $C_{6v}\rightarrow C_{3v}$. A well-studied example
is graphene on semiconducting TMD monolayers in the group-IV family.
The hybridization between $p_{z}$-electrons and the TMD orbitals
generates a spin\textendash valley term $\lambda_{{\rm {sv}}}s_{z}\tau_{z}$
in the continuum model, reflecting the generally different effective
SOC on $A$ and $B$ sublattices \cite{WangMorpurgo2015,MGmitra2016}.
The breaking of sublattice symmetry also generates a mass term $m\thinspace\sigma_{z}$
(of orbital origin), which can exceed tens meV in rotationally-aligned
van der Waals heterostructures \cite{Hunt_14,Jung_15}. Another example
of reduced symmetry occurs in graphene with intercalated $\text{Pb}$
nano-islands \cite{Calleja2015}, where a rectangular superlattice potential leads to an
in-plane spin--valley coupling $\lambda_{{\rm sv}}^{y}\tau_{z}s_{y}$
in Eq.\,(\ref{Hc6v}). Finally, if time-reversal symmetry is broken,
e.g., using a ferromagnetic substrate, a number
of other spin--orbit terms are generally allowed \cite{Phong_17}.

Below, we show that the above picture is further enriched when $\pi$-electrons
in graphene experience a crystal field environment via hybridization to crystal-split states. The interlayer coupling to a low-symmetry substrate removes the rotational invariance from the effective Hamiltonian
Eq.\,(\ref{Hc6v}), leading to a proliferation of\textcolor{black}{{}
}spin\textendash orbit interactions, including in-plane spin--valley
($\lambda_{{\rm sv}}^{y}\tau_{z}s_{y}$) and anisotropic intrinsic-type
($\lambda_{{\rm KM}}^{y}\tau_{z}\sigma_{z}s_{y}$) SOC. To estimate
the strength of the proximity spin\textendash orbit interactions, we consider a minimal tight-binding model for a hybrid
bilayer with hopping parameters obtained from the Slater\textendash Koster
method \cite{SlaterKoster,Harrison}. We present explicit calculations
for two idealized substrates, in which a commensurate monolayer of
heavy atoms sit at hollow and top sites of pristine graphene. Finally,
a Löwdin perturbation scheme is employed to obtain the 
low-energy continuum Hamiltonian. As a concrete example, we then
discuss the possibility of obtaining an enhanced in-plane spin--valley
coupling in a hybrid heterostructure of graphene and a group-IV dichalcogenide
monolayer. The article is organized as follows. In Sec.~\ref{sec1},
we introduce the substrate model and discuss how the eigenstates of
free atomic shells are affected by CFE. In Sec.~\ref{sec2}, we derive
the effective Hamiltonian, when the emergent rotational symmetry
 ($C_{v\infty}$) is broken by the crystal field
environment. In Sec.~\ref{sec3}, we address the scenario where,
added to CFE, the point group $C_{6v}$ is reduced by a sublattice-dependent
interaction with atoms of the substrate, which give rises to new types
of SOCs. In Sec.~\ref{sec4}, we discuss possible realizations with
group-IV TMD monolayers. Section~\ref{sec5} presents our conclusions.

\vspace{0.1in}

\section{Substrate Model\label{sec1} }

\begin{figure}
\vspace{0.1in}
 \centering \includegraphics[width=0.9\columnwidth]{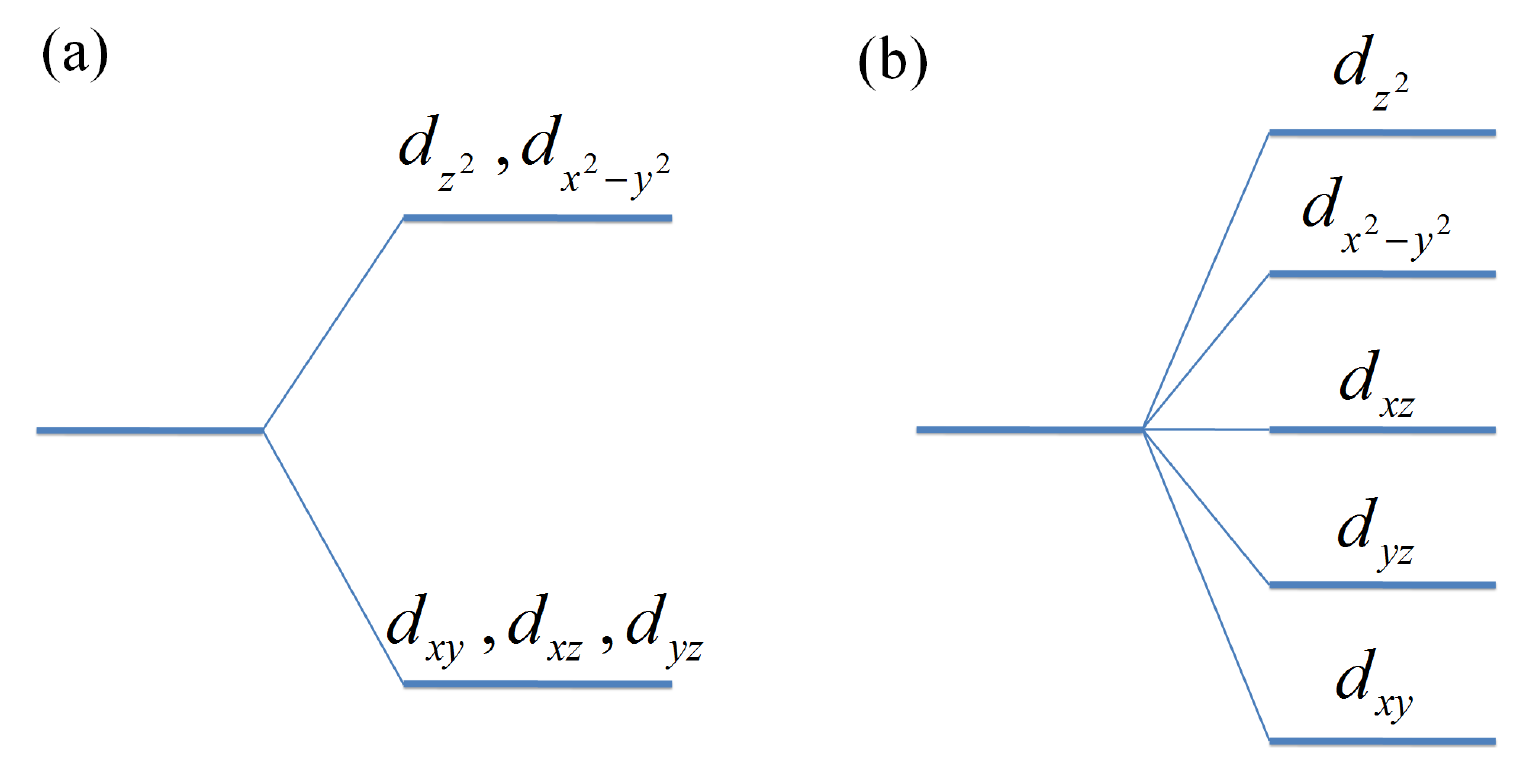} \caption{\label{fig1}Two examples of crystal field splitting. (a) octahedral
(b) orthorhombic.}
\end{figure}

We assume a sufficiently weak interlayer interaction between graphene
and the substrate \cite{MGmitra2016,Phong_17,Giovannetti_07}, so that the electronic states near the Fermi level derive mostly from $p_{z}$- (graphene) states. \textcolor{black}{Since we are
mainly interested in the interplay between CFE and SOC, we shall focus
on substrates containing transition metal atoms. }We focus on atomic
species with outer free shell formed by $d$-states ($l=2$). The
electronic states of a free atom are complex wave functions with well
defined angular momentum projection. When an ion is placed in a crystalline
environment, its electronic states suffer distortions due to the electric
field generated by the surrounding atoms. For $d$ ($l=2$) atomic
states, this effect is usually stronger than the spin\textendash orbit
interaction itself, which can then be treated as a perturbation \cite{FazekasBook}.
The electronic states of a free atom are $(2l+1)$-fold degenerate
(neglecting relativistic corrections), but when the atom is placed in a low-symmetry environment, the
degeneracy is lifted {[}see Fig.\,\ref{fig1}(a){]}. Depending on
the crystal symmetry, some of the original complex atomic
states combine to form real atomic states with no defined angular
momentum projection. If the symmetry is sufficiently low, as in an
orthorhombic crystal, the degeneracy is fully lifted {[}see Fig.\,\ref{fig1}(b){]},
and the atomic wavefunctions are real.

The Hamiltonian is written as $H=H_{g}+H_{\textrm{at}}+V$, where
$H_{g}$ is the standard nearest-neighbor tight-binding Hamiltonian for $\pi$-electrons in graphene and $V$ is the interlayer interaction. To
simplify the analysis, hopping processes within $H_{\textrm{at}}$, as well as disorder effects, are neglected.
Such an approximation suffices for a qualitative description
of the effective (long-wavelength) interactions mediated on graphene \cite{note_local}.
Finally, we assume a general low-symmetry environment, such that the
atomic Hamiltonian for the external free shell subspace reads as\textcolor{black}{{}
$H_{{\rm at}}=H_{0}+H_{{\rm so}}$, with}

\begin{eqnarray}
 &  & H_{0}=\sum_{i}\sum_{s=\uparrow,\downarrow}\sum_{d_{l}}\epsilon_{d_{l}}\big|d_{l},s,i\big>\big<i,s, d_{l}\big|\,,\label{eq:H0orb}\\
 &  & H_{{\rm so}}=\sum_{i}\xi\,\vec{l}_i\cdot\vec{s}_i.\label{eq:H0_SO}\,,
\end{eqnarray}
where $i$ runs over the substrate atoms and $\vec{l}_i$
($\vec{s}_i$) the associated dimensionless orbital (spin) angular momentum
operators. The first term {[}Eq.\,(\ref{eq:H0orb}){]} describes the crystal
field splitting of $d$ levels \cite{comment_energies}. \textcolor{black}{The
second term {[}Eq.}\,\textcolor{black}{(\ref{eq:H0_SO}){]} is the
spin\textendash orbit interaction on the substrate atoms. We note
in passing that CFEs can also lead to anisotropic SOC in Eq.}\,\textcolor{black}{(\ref{eq:H0_SO})
\cite{Santos2017}. Such (usually small) anisotropy is neglected here, since its main effect is simply a modulation of the magnitude of the effective SOCs on graphene.}

We consider two types of commensurate substrates. In the first type,
transition metal atoms of a given species are placed at distance $d$ above the center
of a hexagonal plaquette in graphene (hollow position $h$ in Fig.\,\ref{Figure2}).
In the second type, the atoms are located at a distance $d$ above a carbon
atom (top position). The unit cell of graphene is formed by two sublattices and, as such, there are two possible top configurations:
$t_{A}$ and $t_{B}$ (see Fig.\,\ref{Figure3}). The eigenstates
of the first term Eq.\,(\ref{eq:H0orb}) in space representation
can be written as 
\begin{eqnarray}
\big<\vec{r}\big|d_{l}\big>=R(r)\chi_{l}(\theta,\phi)
\end{eqnarray}
where $R(r)$ is the radial part of wave-function, $\chi_{l}(\theta,\phi)=\big<\theta,\phi\big|d_{l}\big>$,
($l=z^{2},xz,yz,xy,x^{2}-y^{2}$) are tesseral harmonics. Unlike spherical
harmonics (eigenfunctions of $l_{z}$), tesseral harmonics are real
functions and do not have spherical symmetry. For calculation purposes,
we recast the wavefunctions (we omit the radial part hereafter) in
terms of eigenstates of $l_{z}$ as 
\begin{eqnarray}
 &  & \big|d_{z^{2}}\big>=\big|2,0\big>,\label{d1}\\
 &  & \big|d_{xz}\big>=\frac{1}{\sqrt{2}}(-\big|2,1\big>+\big|2,-1\big>),\label{d2}\\
 &  & \big|d_{yz}\big>=\frac{\imath}{\sqrt{2}}(\big|2,1\big>+\big|2,-1\big>),\label{d3}\\
 &  & \big|d_{xy}\big>=\frac{\imath}{\sqrt{2}}(-\big|2,2\big>+\big|2,-2\big>),\label{d4}\\
 &  & \big|d_{x^{2}-y^{2}}\big>=\frac{1}{\sqrt{2}}(\big|2,2\big>+\big|2,-2\big>).\label{d5}
\end{eqnarray}
Below, we show that the main effect of the hybridization of graphene
orbitals with non-spherically symmetric states of the substrate is
to induce anisotropic SOC.

\section{Effective Hamiltonian: hollow position\label{sec2}}

\begin{figure}
\vspace{0.1in}
 \centering \includegraphics[width=0.6\columnwidth]{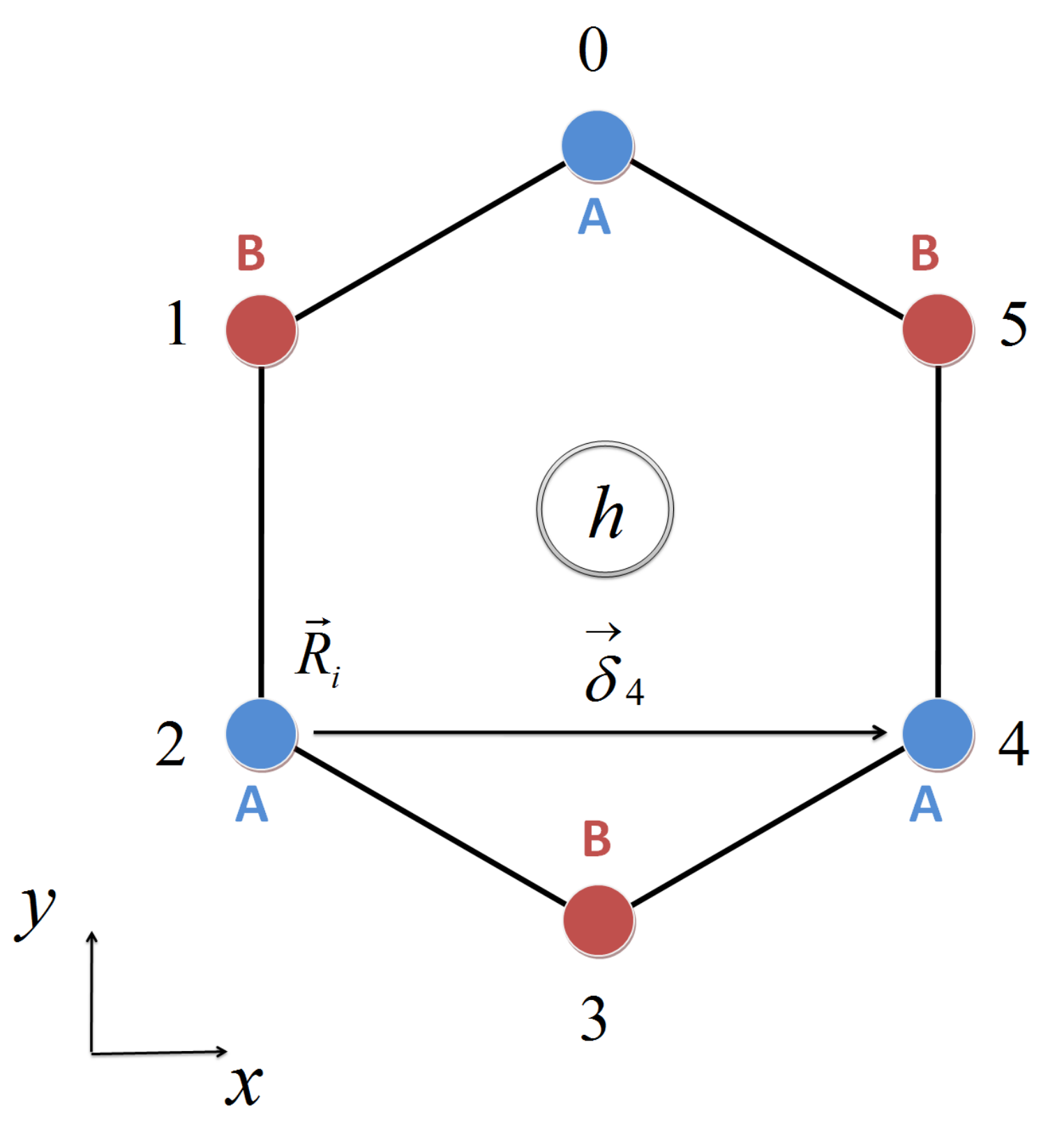} \caption{Hollow-position ($h$). The $j$-index convention is used in Eqs.~(\ref{PhiHollow})
and (\ref{t1h}-\ref{t5h}) and in the definitions of the $\vec{\delta}_{j}$
and the direction cosines $n_{x}^{h},n_{y}^{h},$ and $n_{z}^{h}$.}
\label{Figure2} 
\end{figure}

\begin{figure}
\vspace{0.1in}
 \centering \includegraphics[width=0.8\columnwidth]{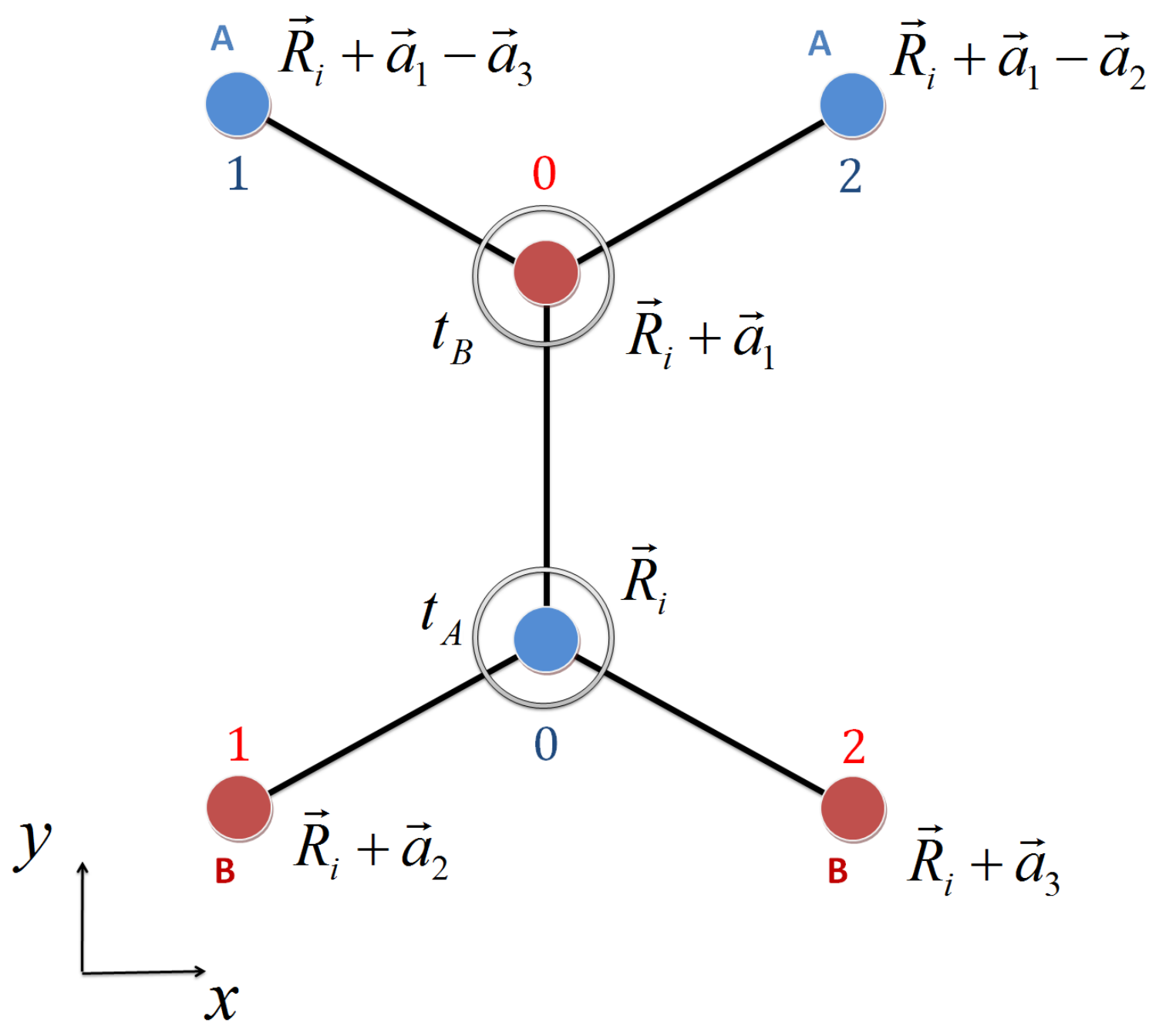} \caption{Unit cell formed by atom $A$ at position $\vec{R}_{i}$ and atom
$B$ at position $\vec{R}_{i}+\vec{a}_{1}$. Atom $t_{A(B)}$ hybridizes
with graphene site $\vec{R}_{i}$ ($\vec{R}_{i}+\vec{a}_1$) (on sublattice $A(B)$) and the
three first neighbouring sites on sublattice $B(A)$. Red (blue) numbers
on $B$($A$) sites define the $j$-index convention used in states
$\big|\Phi_{s,l}^{1A(B)}(\vec{R}_{i})\big>$ and hopping terms of
Eqs~(\ref{t11A}-\ref{t51A}) {[}Eqs.~(\ref{t11B}-\ref{t51B}){]}. }
\label{Figure3} 
\end{figure}

\textcolor{black}{As a simple model for the substrate, we consider a monolayer
of heavy atoms sitting at the hollow sites. The $d$-orbitals
of each substrate atom hybridize with the $p_{z}$ states of the nearest
six carbon atoms (other hoppings are much smaller and thus are neglected).
}The hybridization Hamiltonian is $H_{V}^{{\rm h}}=T_{{\rm h}}+T_{{\rm h}}^{\dagger}$,
with 
\begin{eqnarray}
T_{{\rm h}}=\sum_{\vec{R}_{i}}\sum_{l}\sum_{s=\uparrow,\downarrow}\big|\Phi_{s,l}(\vec{R}_{i})\big>\big<d_{l},s,\vec{R}_{i}+\vec{h}\big|,\label{Thollow}
\end{eqnarray}
where $\vec{R}_{i}$ are lattice vectors, $\vec{h}$ is the position of $h$ inside the plaquette, $s=\pm1$ for (up)
down states and 
\begin{eqnarray}
\big|\Phi_{s,l}(\vec{R}_{i})\big>=\sum_{j=0}^{5}t_{l,s,j}\big|\sigma_{j},s,\vec{R}_{i}+\vec{\delta}_{j}\big>.\label{PhiHollow}
\end{eqnarray}
Here, $j=0,...,5$ runs anti-clockwise and follows the convention
in Fig.~\ref{Figure2} and $\sigma_{j}=A(B)$ for even (odd) $j$.
The substrate\textendash graphene hopping amplitudes are defined by
$t_{l,s,j}=\big<\sigma_{j},s,\vec{R}_{i}+\vec{\delta}_{j}\big|\hat{V}\big|d_{l},s,\vec{R}_{i}+\vec{h}\big>$, 
where $\vec{\delta}_{j}$ are vectors connecting neighboring carbon atoms \cite{ReviewNuno}; see Fig.~\ref{Figure2}.

The hopping amplitudes are evaluated by means of the Slater\textendash Koster
approach \cite{SlaterKoster,Harrison} 
\begin{eqnarray}
 &  & \big<p_{z}\big|V\big|d_{xy}\big>=n_{x}n_{y}n_{z}(\sqrt{3}V_{pd\sigma}-2V_{pd\pi}),\label{sk1}\\
 &  & \big<p_{z}\big|V\big|d_{x^{2}-y^{2}}\big>=\frac{\sqrt{3}}{2}n_{z}(n_{x}^{2}-n_{y}^{2})V_{pd\sigma}-n_{z}(n_{x}^{2}-n_{y}^{2})V_{pd\pi},\nonumber \\
\label{sk2}\\
 &  & \big<p_{z}\big|V\big|d_{zx}\big>=\sqrt{3}n_{z}^{2}n_{x}V_{pd\sigma}+(1-2n_{z}^{2})n_{x}V_{pd\pi},\ \label{sk3}\\
 &  & \big<p_{z}\big|V\big|d_{zy}\big>=\sqrt{3}n_{z}^{2}n_{y}V_{pd\sigma}+(1-2n_{z}^{2})n_{y}V_{pd\pi},\ \label{sk4}\\
 &  & \big<p_{z}\big|V\big|d_{z^{2}}\big>=\sqrt{3}n_{z}(n_{x}^{2}+n_{y}^{2})V_{pd\pi}\nonumber \\
 &  & ~~~~~~~~~~~~~~~~~-\frac{1}{2}n_{z}(n_{x}^{2}+n_{y}^{2}-2n_{z}^{2})V_{pd\sigma},\label{sk5}
\end{eqnarray}
where $V_{pd\sigma}$ and $V_{pd\pi}$ are two-centers integrals,
which can be obtained by quantum chemistry methods or by fitting to
first-principles electronic structure calculations \cite{Cappelluti2013,SilvaGuillen2016}.
$n_{i}$ are direction cosines of the vector connecting a $j$-carbon
atom and the substrate atom at $h$ position. The hopping amplitudes
are given in the Appendix.

We are interested in the low-energy theory near the Dirac points $\vec{K}=-\vec{K'}=\frac{4\pi}{3a}\hat{x}$.
The Fourier transform of the hopping matrix at these points can be
easily computed and we obtain, for each valley ($\tau = \pm 1$) \begin{widetext}

\begin{eqnarray}
 &  & T_{{\rm h}}=\sum_{s=\uparrow,\downarrow}\imath\tau\frac{3V_{1}}{\sqrt{2}}e^{\imath\tau2\pi/3}\big(\big|A,s\big\rangle+\big|B,s\big\rangle\big)\big<d_{xz},s\big|+\frac{3V_{1}}{\sqrt{2}}e^{\imath\tau2\pi/3}\big(\big|A,s\big>-\big|B,s\big>\big)\big<d_{yz},s\big|\nonumber \\
 &  & +\imath\tau\frac{3V_{2}}{\sqrt{2}}e^{\imath\tau2\pi/3}\big(\big|B,s\big>-\big|A,s\big>\big)\big<d_{xy},s\big|-\frac{3V_{2}}{\sqrt{2}}e^{\imath\tau2\pi/3}\big(\big|B,s\big>+\big|A,s\big>\big)\big<d_{x^{2}-y^{2}},s\big|\,\,,\label{Td}
\end{eqnarray}
\end{widetext}The various constants read as $V_{0}=\sqrt{3}n(1-n^{2})V_{pd\pi}-\frac{1}{2}n(1-3n^{2})V_{pd\sigma}$,
$V_{1}=\frac{1}{\sqrt{2}}\sqrt{1-n^{2}}(\sqrt{3}n^{2}V_{pd\sigma}+(1-2n^{2})V_{pd\pi})$,
and $V_{2}=\frac{1}{\sqrt{2}}n(1-n^{2})\big(\sqrt{3}V_{pd\sigma}/2-V_{pd\pi}\big)$,
where $n_{z}^{h}=n=a_{0}/\sqrt{a_{0}^{2}+d^{2}}$, with $a_{0}$ being
the distance between two carbon atoms.

Next, we use degenerate perturbation theory to obtain a graphene-only
effective Hamiltonian 
\begin{eqnarray}
H_{{\rm eff}}^{{\rm h}} & = & -T_{{\rm h}}(H_{0}+H_{\textrm{so}})^{-1}T_{{\rm h}}^{\dagger}\nonumber \\
 & \approx & -T_{{\rm h}}H_{0}^{-1}T_{{\rm h}}^{\dagger}+T_{{\rm h}}H_{0}^{-1}H_{\textrm{so}}H_{0}^{-1}T_{{\rm h}}^{\dagger}\,,\label{heff}
\end{eqnarray}
where we treated the spin\textendash orbit term of the substrate Hamiltonian
$H_{{\rm so}}$ as a next-order perturbation compared to $H_{0}$.
The first term $H_{{\rm h}}^{{\rm CF}}=-T_{{\rm h}}H_{0}^{-1}T_{{\rm h}}^{\dagger}$
can be expressed in terms of Pauli matrices: 
\begin{eqnarray}
H_{{\rm h}}^{{\rm CF}} & = & -\lambda_{0}-\lambda_{x}\sigma_{x},\label{hcf}
\end{eqnarray}
with
\begin{eqnarray}
 &  & \lambda_{0(x)}=\frac{9(V_{1})^{2}}{2\epsilon_{xz}}\pm\frac{9(V_{1})^{2}}{2\epsilon_{yz}}+\frac{9(V_{2})^{2}}{2\epsilon_{xy}}\pm\frac{9(V_{2})^{2}}{2\epsilon_{x^{2}-y^{2}}},\label{l0h}
\end{eqnarray}
The first term in Eq.~(\ref{hcf}) is a trivial energy shift. The
interaction $\lambda_{x}$ is an orbital term, which can be absorbed
by a redefinition of $k_{x}$ in Eq.~(\ref{Hc6v}). The interplay
between SOC and CFE is captured by the second term, $H_{{\rm h}}^{{\rm CF/SO}}=T_{{\rm h}}H_{0}^{-1}H_{so}H_{0}^{-1}T_{{\rm h}}^{\dagger}$,
which has the form 
\begin{eqnarray}
H_{{\rm h}}^{{\rm CF/SO}} & = & -\lambda_{{\rm R}}^{1}\sigma_{y}s_{x}-\lambda_{{\rm R}}^{2}\tau_{z}\sigma_{x}s_{y}\nonumber \\
 &  & +\lambda_{{\rm KM}}\tau_{z}\sigma_{z}s_{z}+\lambda_{{\rm sv}}^{y}\tau_{z}s_{y},\label{hcfso}
\end{eqnarray}
with couplings determined by 
\begin{eqnarray}
 &  & \lambda_{{\rm R}}^{1}=18\xi V_{1}V_{2}\Big(\frac{1}{\epsilon_{xy}\epsilon_{xz}}+\frac{1}{\epsilon_{yz}\epsilon_{x^{2}-y^{2}}}\Big),\label{lra}\\
 &  & \lambda_{{\rm R}}^{2}=18\xi V_{1}V_{2}\Big(\frac{1}{\epsilon_{xy}\epsilon_{yz}}+\frac{1}{\epsilon_{xz}\epsilon_{x^{2}-y^{2}}}\Big),\label{lrb}\\
 &  & \lambda_{{\rm KM}}=9\xi\Big(\frac{(V_{1})^{2}}{\epsilon_{yz}\epsilon_{xz}}-\frac{2(V_{2})^{2}}{\epsilon_{xy}\epsilon_{x^{2}-y^{2}}}\Big),\label{lkm}\\
 &  & \lambda_{{\rm sv}}^{y}=9\xi V_{1}V_{2}\Big(\frac{1}{\epsilon_{xy}\epsilon_{yz}}-\frac{1}{\epsilon_{xz}\epsilon_{x^{2}-y^{2}}}\Big).\label{lsvy}
\end{eqnarray}
The first two terms in Eq.~(\ref{hcfso}) form an anisotropic Bychkov-Rashba
coupling. The third term is the familiar intrinsic-like SOC. The last
term is an in-plane spin\textendash valley coupling, leading to an
anisotropic spectrum. \textcolor{black}{Note that this term vanishes
in the absence of crystal field splitting. The same effective couplings
in Eqs.~(\ref{hcf}) and (\ref{hcfso}) were obtained in Ref.~\cite{Calleja2015}
for $\text{Pb}$ atoms in the absence of CFE due to the reduced point
group symmetry $C_{2v}$ of the underlying superlattice. }\vspace{0.1in}

\section{Effective Hamiltonian: Top Position\label{sec3}}

We assume that the top positions ($t_{A}$ and $t_{B}$) are occupied
by different atomic species (or, equivalently, equal species placed
at different distances from graphene). This accounts for the important
class of a graphene interface with reduced point group symmetry $C_{3v}$
(in the absence of CFE). Such a sublattice-dependent interaction was
absent in the hollow position case. The hybridization between $p_{z}$
orbitals of graphene and $d$-atoms on top position can be written
as $H_{V}=T_{{\rm t}}+T_{{\rm t}}^{\dagger}$, where the hopping matrix
$T_{t}$ is given by 

\begin{eqnarray}
T_{{\rm t}} &  & =\sum_{\vec{R}_{i}}\sum_{l}\sum_{s=\uparrow,\downarrow}\big(\big|\Phi_{s,l}^{(0,A)}(\vec{R}_{i})\big>+\big|\Phi_{s,l}^{(1,A)}(\vec{R}_{i})\big>\big)\big<d_{l},s,A,\vec{R}_{i}\big|\nonumber \\
 &  & +\big(\big|\Phi_{s,l}^{(0,B)}(\vec{R}_{i})\big>+\big|\Phi_{s,l}^{(1,B)}(\vec{R}_{i})\big>\big)\big<d_{l},s,B,\vec{R}_{i}+\vec{a}_{1}\big|\,,\label{Tt}
\end{eqnarray}
and the $\Phi$-states are defined in similar way to the hollow position
case, namely, 
\begin{eqnarray}
 &  & \big|\Phi_{s,l}^{(0,A)}(\vec{R}_{i})\big>=t_{l,s}^{(0,A)}\big|A,s,\vec{R}_{i}\big>,\label{phi0A}\\
 &  & \big|\Phi_{s,l}^{(1,A)}(\vec{R}_{i})\big>=\sum_{j=0}^{2}t_{l,s,j}^{(1,A)}\big|B,s,\vec{R}_{i}+\vec{a}_{j+1}\big>,\label{phi1A}
\end{eqnarray}
\textcolor{black}{where $t_{l,s}^{(0,A)}=\big<A,s,\vec{R}_{i}|V\big|d_{l},s,A,\vec{R}_{i}\big>$
and $t_{l,s,j}^{(1,A)}=\big<B,s,\vec{R}_{i}+\vec{a}_{j+1}\big|V\big|d_{l},s,A,\vec{R}_{i}\big>$.
Similar definitions are employed for states $\big|\Phi_{s,l}^{(0/1,A/B)}(\vec{R}_{i})\big>$
in Eq.~(\ref{Tt}).}

The hopping parameters are written in the Appendix. Around $K$ points
in the hexagonal Brillouin zone, the hopping matrix can be written
as $T_{{\rm t}}=\sum_{\tau=\pm1}T_{A{\rm t}}+T_{B{\rm t}}$ where\begin{widetext}
\begin{widetext}
\begin{eqnarray}
T_{{\rm t}}^{A(B)}= &  & \sum_{s=\uparrow,\downarrow}V_{0}^{0A(B)}\big|A(B),s\big>\big<d_{z^{2}},s,A(B)\big|\pm\frac{3V_{1}^{1A(B)}}{\sqrt{2}}\big|B(A),s\big>\big<d_{yz},s,A(B)\big|+\imath\tau\frac{3V_{1}^{1A(B)}}{\sqrt{2}}\big|B(A),s\big>\big<d_{xz},s,A(B)\big|\nonumber \\
 &  & -\frac{3V_{2}^{1A(B)}}{\sqrt{2}}\big|B(A),s\big>\big<d_{x^{2}-y^{2}},s,A(B)\big|\mp\imath\tau\frac{3V_{2}^{1A(B)}}{\sqrt{2}}\big|B(A),s\big>\big<d_{xy},s,A(B)\big|,\label{tB}
\end{eqnarray}
\end{widetext}with the various constants given in Appendix. Degenerate
perturbation theory yields 
\begin{eqnarray}
H_{{\rm t}}^{{\rm CF}} &  & =-T_{{\rm t}}^{A}H_{0}^{-1}{T_{{\rm t}}^{A}}^{\dagger}-T_{{\rm t}}^{B}H_{0}^{-1}{T_{{\rm t}}^{B}}^{\dagger}\nonumber \\
 &  & =-\tilde{\lambda}_{0}-\Delta\sigma_{z}.\label{hcft}
\end{eqnarray}
with coupling constants 
\begin{eqnarray}
\tilde{\lambda}_{0} & = & \,\,\,\frac{(V_{0}^{0A})^{2}}{2\epsilon_{z^{2}}^{A}}+\frac{9(V_{1}^{1A})^{2}}{4\epsilon_{xz}^{A}}+\frac{9(V_{1}^{1A})^{2}}{4\epsilon_{yz}^{A}}+\frac{9(V_{2}^{1A})^{2}}{4\epsilon_{x^{2}-y^{2}}^{A}}\nonumber \\
 &  & +\,\,\frac{9(V_{2}^{1A})^{2}}{4\epsilon_{xy}^{A}}+(A\rightarrow B),\label{l0A}\\
\Delta & = & \,\,\,\frac{(V_{0}^{0A})^{2}}{2\epsilon_{z^{2}}^{A}}-\frac{9(V_{1}^{1A})^{2}}{4\epsilon_{xz}^{A}}-\frac{9(V_{1}^{1A})^{2}}{4\epsilon_{yz}^{A}}-\frac{9(V_{2}^{1A})^{2}}{4\epsilon_{x^{2}-y^{2}}^{A}}\nonumber \\
 &  & \,\,-\frac{9(V_{2}^{1A})^{2}}{4\epsilon_{xy}^{A}}-(A\rightarrow B),\label{DeltaAB}
\end{eqnarray}
representing an energy shift and a staggered sublattice potential, respectively.
The combined effect of crystal field and SOC can be written as 
\begin{eqnarray}
H_{{\rm t}}^{{\rm CF/SO}} & = & T_{{\rm t}}^{A}H_{0}^{-1}H_{\textrm{so}}H_{0}^{-1}{T_{{\rm t}}^{A}}^{\dagger}+T_{{\rm t}}^{B}H_{0}^{-1}H_{\textrm{so}}H_{0}^{-1}{T_{{\rm t}}^{B}}^{\dagger}\nonumber \\
 & = & -\tilde{\lambda}_{{\rm R}}^{1}\sigma_{y}s_{x}-\tilde{\lambda}_{{\rm R}}^{2}\tau_{z}\sigma_{x}s_{y}+\tilde{\lambda}_{{\rm KM}}\tau_{z}\sigma_{z}s_{z}\nonumber \\
 &  & +\tilde{\lambda}_{{\rm sv}}^{z}\tau_{z}s_{z}+\tilde{\lambda}_{{\rm KM}}^{y}\tau_{z}\sigma_{z}s_{y}+\tilde{\lambda}_{{\rm sv}}^{y}\tau_{z}s_{y},\label{HSOt}
\end{eqnarray}
where the coupling constants are given by 
\begin{eqnarray}
 &  & \tilde{\lambda}_{{\rm R}}^{1}=3\sqrt{6}\xi\left[\frac{V_{0}^{0A}V_{1}^{1A}}{\epsilon_{yz}^{A}\epsilon_{z^{2}}^{A}}+(A\rightarrow B)\right],\label{tR1}\\
 &  & \tilde{\lambda}_{{\rm R}}^{2}=3\sqrt{6}\xi\left[\frac{V_{0}^{0A}V_{1}^{1A}}{\epsilon_{xz}^{A}\epsilon_{z^{2}}^{A}}+(A\rightarrow B)\right],\label{tR2}\\
 &  & \tilde{\lambda}_{{\rm KM}}=-{9\xi}\left[\frac{(V_{1}^{1A})^{2}}{2\epsilon_{yz}^{A}\epsilon_{xz}^{A}}-\frac{(V_{2}^{1A})^{2}}{\epsilon_{x^{2}-y^{2}}^{A}\epsilon_{xy}^{A}}+(A\rightarrow B)\right],\label{tKM}\\
 &  & \tilde{\lambda}_{{\rm sv}}^{z}={9\xi}\left[\frac{(V_{1}^{1A})^{2}}{2\epsilon_{yz}^{A}\epsilon_{xz}^{A}}-\frac{(V_{2}^{1A})^{2}}{\epsilon_{x^{2}-y^{2}}^{A}\epsilon_{xy}^{A}}-(A\rightarrow B)\right],\label{tsvz}\\
 &  & \tilde{\lambda}_{{\rm KM}}^{y}=-\frac{9}{4}\xi\left[\frac{V_{1}^{1A}V_{2}^{1A}}{\epsilon_{xy}^{A}\epsilon_{yz}^{A}}-\frac{V_{1}^{1A}V_{2}^{1A}}{\epsilon_{x^{2}-y^{2}}^{A}\epsilon_{xz}^{A}}-(A\rightarrow B)\right],\label{tkmy}\\
 &  & \tilde{\lambda}_{{\rm sv}}^{y}=\frac{9}{4}\xi\left[\frac{V_{1}^{1A}V_{2}^{1A}}{\epsilon_{xy}^{A}\epsilon_{yz}^{A}}-\frac{V_{1}^{1A}V_{2}^{1A}}{\epsilon_{x^{2}-y^{2}}^{A}\epsilon_{xz}^{A}}+(A\rightarrow B)\right].\label{tsvy}
\end{eqnarray}
\end{widetext}

In addition to the SOCs already obtained in the hollow case, the combination
of a sublattice dependent interaction and CFE gives rise to new terms.
We obtain the expected spin\textendash valley coupling $\tilde{\lambda}_{{\rm sv}}^{z}\tau_{z}s_{z}$,
which together with the Bychkov-Rashba SOC are the dominant spin\textendash orbit
interactions in group-VI TMD\textendash graphene bilayers~\cite{WangMorpurgo2015,MGmitra2016}.
Interestingly, the broken orbital degeneracy in the substrate also
generates an in-plane intrinsic spin-orbit coupling $\tilde{\lambda}_{{\rm KM}}^{y}\tau_{z}\sigma_{z}s_{y}$.
This term can open a quantum spin Hall insulating gap that is robust
against Bychkov-Rashba SOC. A more detailed analysis of the effect
of this interaction will be given in the next section.

It is instructive to consider two different limiting cases. First
we consider the situation where all the energies $d$ orbitals of
the substrate are degenerate, i.e., absence of a CFE. By analyzing
the coupling constants on equations (\ref{tR1}-\ref{tsvy}) the only
couplings that remain are the familiar isotropic Bychkov-Rashba coupling,
intrinsic-like SOC and the spin--valley term. These same couplings
were obtained for TMD/graphene heterostructures \cite{WangMorpurgo2015,MGmitra2016}
and enable interesting spin-dependent phenomena, such as anisotropic
spin lifetime \cite{Cummings2017}, spin Hall effect \cite{Milletari2017}
and inverse spin galvanic effect \cite{Offidani2017}. The second
limit case is when top positions $t_{A}$ and $t_{B}$ are equivalent,
so that one has $V_{0}^{0A}=V_{0}^{0B}$, $V_{1}^{1A}=V_{1}^{1B}$
and $V_{2}^{1A}=V_{2}^{1B}$. For this situation the SOCs that appears
are the same of Eq. (\ref{hcfso}) for hollow position case due to
the restoration of sublattice symmetry.

\section{Discussion \label{sec4}}

This paper aims to explore the modifications to the electronic states of graphene placed on a substrates characterized by a crystal field
environment. In a realistic scenario, we expect the proximity-induced SOC to be sensitive
to the type of crystal field splitting and the valence
of the substrate atoms. A quantitative analysis is beyond the scope of
this work. Nevertheless, the crystal field is expected to be significant
in compounds containing transition metals atoms, in which the incomplete
outer shell is formed by $d$ electrons. The electronic structure
of certain TMDs is known to be strongly affected by CFE on the atomic states of
transition metal (TM) atoms \cite{Gong2013,Zhang2015}. TMD layers
consist of an hexagonally close-packed sheet of TM atoms sandwiched
between sheets of chalcogen atoms and their metal coordination can
be either trigonal prismatic or octahedral. In the trigonal prismatic
coordination, the two chalcogen sheets are stacked directly above each
other (known as H phase). The stacking order in the octahedral phase
(T phase) is ABC and the chalcogen atoms of one of the sheets can
be located at the center of the honeycomb lattice. In this case, the
coordination of the TM atoms is octahedral.

Group IV TMDs have an octahedral structure, whereas group VI TMDs,
of the well studied W and Mo compounds, tend to display a trigonal
prismatic geometry, whereas both octahedral and trigonal prismatic
phases are observed in group V TMDs. The \emph{trigonal prismatic
geometry} enforces a splitting of $d$ orbitals in an single state
$d_{z^{2}}$ and two doublets $d_{x^{2}-y^{2}}/d_{xy}$ and $d_{xz}/d_{yz}$.
On the other hand, in the \emph{octahedral} geometry, a doublet $d_{z^{2}}/d_{x^{2}-y^{2}}$
and a triplet $d_{xy}/d_{xz}/d_{yz}$ are formed. Going back to Eqs.~(\ref{hcfso})
and (\ref{HSOt}), one can see that the main signatures of the CFE
is the broken rotational symmetry in the continuum due to the hybridization
of graphene with states without spherical symmetry. The latter results
in a in-plane spin--valley coupling, $\lambda_{{\rm {sv}}}^{y}\tau_{z}s_{y}$
and an anisotropic Bychkov-Rashba SOC. For both top- and hollow- position
cases, it is necessary that $\epsilon_{xy}\epsilon_{yz}\neq\epsilon_{x^{2}-y^{2}}\epsilon_{xz}$,
for the appearance of the in-plane spin--valley coupling, which is
the case of TM atoms with an \emph{octahedral} distortion; see Fig.~\ref{fig1}(a).
This type of crystal field is found in the group IV family ($XY_{2}$
where $X=\text{Zr,Hf}$, and $Y=\text{S,Se,Te}$) and opens up the
possibility to observe this coupling in bilayers of these materials
and graphene. Less attention has been paid to this family \cite{Jiang2011,Li2014}
compared to group V and VI TMDs. The application of $\text{Zr}$-based
chalcogenides in solar energy devices has been suggested \cite{Jiang2011},
and the possibility of tuning its properties by pressure, electric
field and phase engineering was recently explored in density functional
theory calculations \cite{Kumar2015}. Our findings suggest that TMDs
of family IV are potential candidates to induce non-trivial spin textures
in graphene via proximity coupling. On the other hand, the anisotropic
Bychkov-Rashba coupling requires $\epsilon_{xz}\neq\epsilon_{yz}$,
which is only possible in a very low symmetry environment. The low
symmetric $\text{T}'$-phase in WTe$_{2}$ monolayers, that presents
a quantum spin Hall phase \cite{Qian2014,Wu2018}, could induce an
anisotropic Rashba coupling in graphene. This type of anisotropy can
lead to an increase of the spin Hall angle in graphene decorated with
SOC impurities \cite{Cazalilla2016}.

\begin{figure}
\vspace{0.1in}
 \centering \includegraphics[width=0.6\columnwidth]{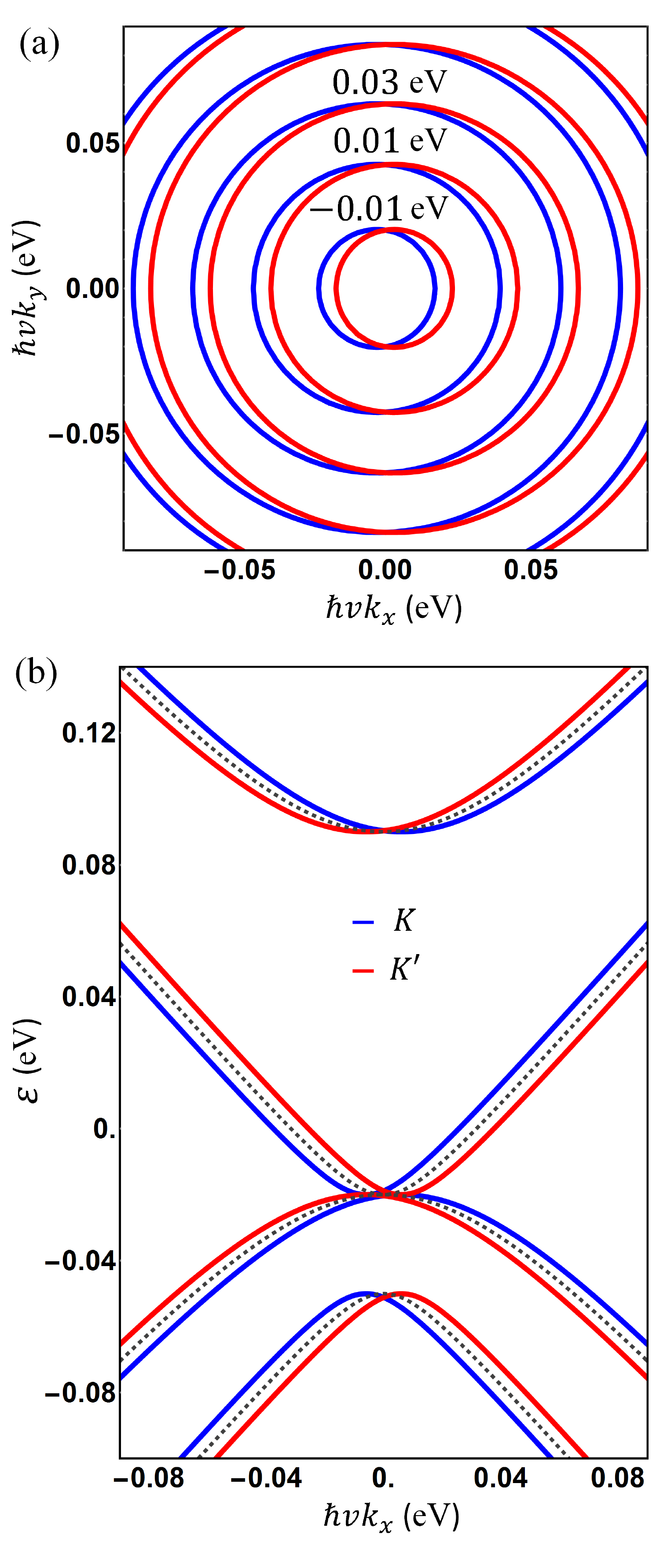} \caption{(a) Fermi surface contours around $K$($K'$) points.
(b) Low-energy spectrum along $k_{x}$ direction ($k_y=0$). Parameters: $\lambda_{\textrm{sv}}^{y}=6\text{meV}$,
$\lambda_{\textrm{R}}^{1}=\lambda_{\textrm{R}}^{2}=35\text{meV}$,
and $\lambda_{\textrm{KM}}=20\text{meV}$.}
\label{Figure5} 
\end{figure}

For large interlayer distances,\textcolor{black}{{}
the overlap matrix between states centred on different
atomic positions can be neglected, a}nd we can use Eqs.~(\ref{lra})-(\ref{lsvy}) and (\ref{tR1})-(\ref{tsvy})
to perform a rough estimative of the different SOCs. Using Slater\textendash Koster
parameters for TM-carbon bonds as reported in Ref.~\cite{Le1991}
and the crystal-field splitting and spin\textendash orbit energy ($\xi$)
reported in Ref.~\onlinecite{Jiang2011}, we estimate the graphene
effective SOCs for distances $\approx5$ times the graphene's
lattice spacing. The dominant SOCs are found to be Kane--Mele and Rashba couplings,
with estimated magnitude in the range $20-40\text{meV}$ for both
hollow and top substrate atoms, which is consistent with the
robust weak antilocalization features in magnetocondutance measurements
\cite{Whang2017}. The in-plane spin--valley SOC $\lambda_{{\rm sv}}^{y}$
is one order of magnitude weaker, being $\approx2.5\text{meV}$ for
the hollow position case, and $\approx1.2\text{meV}$ for the top
position case (when atoms $A$ and $B$ have the same nature), which
suggests a small but observable effect. For short
graphene-substrate separations, numerical estimations need to take
into account the overlap matrix between states at different atomic
positions, which is beyond the scope of the present work. Note that the interlayer
distance can be tuned by external pressure \cite{Kumar2015}, which
can be employed to tailor the SOC. Figure~\ref{Figure5} shows the low-energy spectrum along $k_{x}$-direction
when graphene has an effective SOC formed by Rashba, Kane-Mele and
in-plane spin--valley interactions. We see a interesting feature on
this spectrum: the energy dispersion around inequivalent valleys is
shifted (along $k_{x}$-direction) with respect to the bare graphene
Dirac spectrum. This shift has opposite signs at inequivalent valleys
as required by time-reversal symmetry.

\begin{figure}
\vspace{0.1in}
 \centering \includegraphics[scale=0.55]{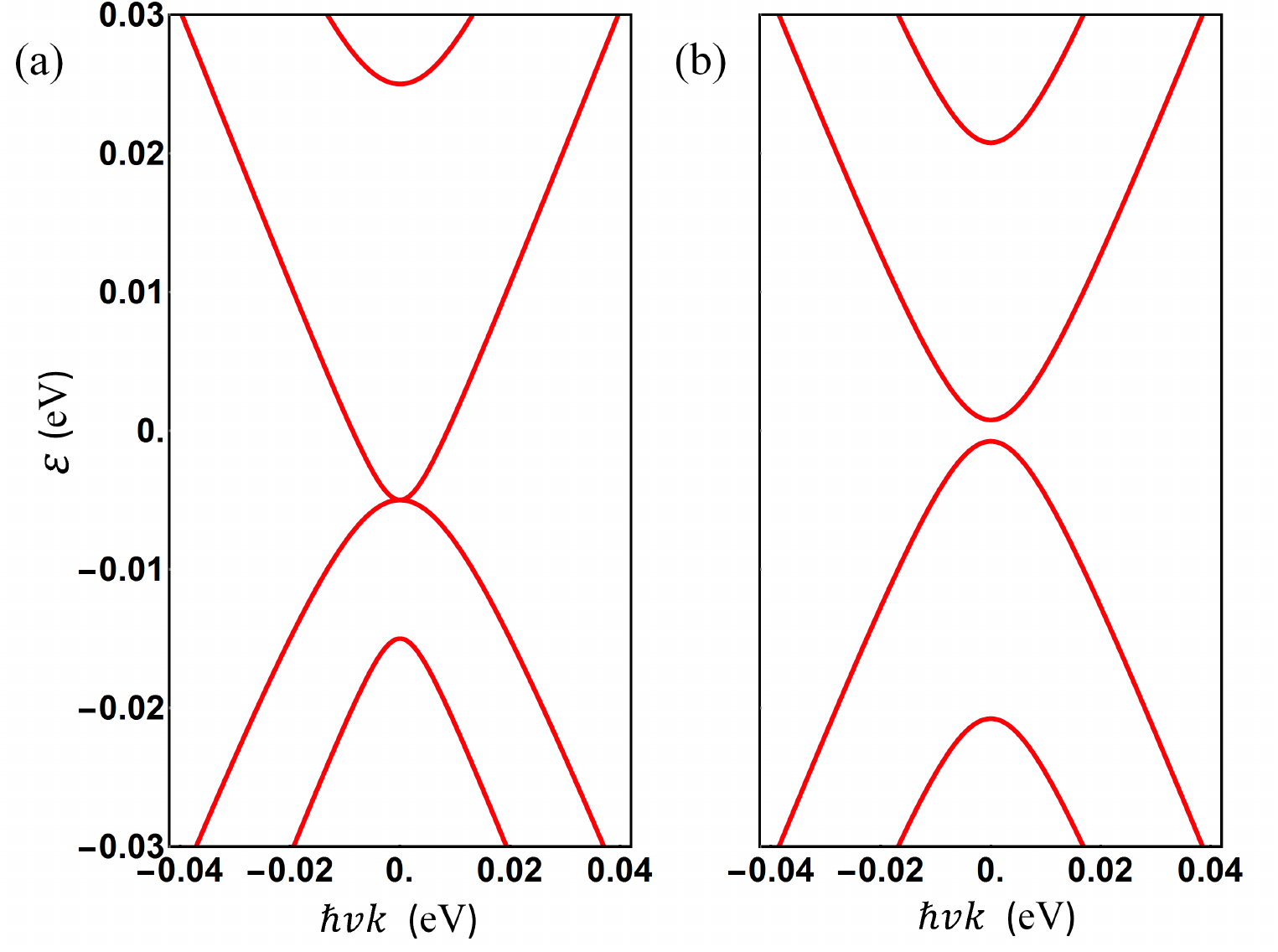} \caption{Energy spectrum of graphene placed on a high SOC substrate with a
crystal field environment. (a) $\lambda_{{\rm R}}=10\,\text{meV}$,
$\lambda_{{\rm KM}}=5\,\text{meV}$; (b) $\lambda_{{\rm R}}=10\,\text{meV}$,
$\lambda_{{\rm KM}}^{y}=5\,\text{meV}$. The gap has a nontrivial
$\mathbb{Z}_{2}$ topological character corresponding to a quantum
spin Hall phase.}
\label{Figure6} 
\end{figure}

Finally, we discuss the in-plane spin\textendash orbit interaction
$\lambda_{{\rm KM}}^{y}\tau_{z}\sigma_{z}s_{y}$ in Eq.~(\ref{HSOt}).
In our estimate for group IV-TMD\textendash graphene bilayers this
type of coupling is relatively weak, being of the same order of the
in-plane spin--valley term ($\approx$ 1 meV). However, it has interesting
topological properties. As mentioned above, this SOC can induce a nontrivial
topological insulating gap associated to a spin Chern number $C=2$ \cite{Sheng_06}.
However, the robustness of the $\mathbb{Z}_{2}$ topological phase
differs from that generated by the familiar intrinsic SOC in graphene
$\lambda_{{\rm KM}}$ \cite{KaneMele1}. When only $C_{6v}$-invariant
SOCs are present, that is $\lambda_{{\rm KM}}$ and $\lambda_{{\rm R}}$,
the quantum spin Hall gap closes if $|\lambda_{{\rm R}}|>|\lambda_{{\rm KM}}|$
\cite{KaneMele1}, destroying the topological phase; see Fig. \ref{Figure6}
(a). On the other hand, if the $\mathbb{Z}_{2}$ topological phase
is a consequence of $\lambda_{{\rm KM}}^{y}$ , the gap remains finite
for \textit{any} value of $\lambda_{{\rm R}}$ and as long as $|\lambda_{\textrm{sv}}^{y}|<|\lambda_{\textrm{KM}}^{y}|$.
A typical band structure is shown in Fig.\,\ref{Figure6}(b), where
the topological gap is finite even for large Bychkov-Rashba coupling.
$\lambda_{{\rm R}}$ is one of the main obstacles to the observation
of quantum spin Hall effect in graphene because of its interplay with
$\lambda_{{\rm KM}}$. Our analysis suggest that realistic hybrid
graphene\textendash TMD bilayers can host a novel type of quantum
spin Hall insulator with fully in-plane helical edge states. \vspace{0.1in}

\section{Conclusion\label{sec5}}

We studied theoretically proximity spin\textendash orbital effects
in graphene placed on low-symmetry substrates with broken orbital degeneracy.
We derived a low-energy (long-wavelength) theory for an idealized monolayer substrate,
which allowed us to demonstrate a simple mechanism to remove the rotational invariance of electronic states in proximity-coupled
graphene, i.e., their hybridization to crystal-field split states.
The low symmetry environment was shown to render spin\textendash orbit
interactions of $\pi$-electrons highly anisotropic. The most distinctive
signature of the crystal field effect is the appearance of in-plane
Zeeman 'spin--valley' interaction $\lambda_{{\rm {sv}}}$ and anisotropic
intrinsic-type spin-orbit coupling $\lambda_{{\rm KM}}^{y}$, which
can drive a transition to a quantum spin Hall insulating phase displaying in-plane helical edge states. As
a possible candidate to observe the predicted effects, we suggested
group IV TMD monolayers, where transition metal atoms have an octahedral
distortion and contain the necessary ingredients to induce anisotropic
in-plane SOCs on graphene.

\section*{Acknowledgments}

T. G. R. acknowledges the support from the Newton Fund and the Royal
Society (U.K.) through the Newton Advanced Fellowship scheme (ref.
NA150043), T. P. C. and T. G. R. thank the Brazilian Agency CNPq for
financial support. A.F. gratefully acknowledges the support from the
Royal Society (U.K.) through a Royal Society University Research Fellowship.

\section*{Appendix \label{sec:Appendix.}}

\subsection{Hopping parameters and $\Phi$-states. \label{AppendixA}}

The explicit expressions for the hopping parameters in Eq.~(\ref{PhiHollow})
for the hollow-position case are
\begin{eqnarray}
 &  & t_{z^{2},s,j}=V_{0},\label{t1h}\\
 &  & t_{xz,s,j}=i\frac{V_{1}}{\sqrt{2}}(e^{i\pi j/3}-e^{-i\pi j/3}),\label{t2h}\\
 &  & t_{yz,s,j}=\frac{V_{1}}{\sqrt{2}}(e^{i\pi j/3}+e^{-i\pi j/3}),\label{t3h}\\
 &  & t_{xy,s,j}=i\frac{V_{2}}{\sqrt{2}}(e^{i2\pi j/3}-e^{-i2\pi j/3}),\label{t4h}\\
 &  & t_{x^{2}-y^{2},s,j}=-\frac{V_{2}}{\sqrt{2}}(e^{i2\pi j/3}+e^{-i2\pi j/3}),\label{t5h}
\end{eqnarray}
with constants $V_{0}$, $V_{1}$ and $V_{2}$ given in Sec.~\ref{sec2}.
$\Phi$-states of Eq. (\ref{PhiHollow}) can be written in terms of
hexagonal states
\begin{eqnarray}
\big|\Omega_{m}^{s}(\vec{R}_{i})\big>=\sum_{j=0}^{5}e^{im\pi j/3}\big|\sigma_{j},s,\vec{R}_{i}+\vec{\delta}_{j}\big>.\label{hexagonal}
\end{eqnarray}
with well-defined angular momentum $l_{z}=\hbar m$  and are described
in reference \cite{Pachoud2014}. Using Eqs. (\ref{t1h}-\ref{t5h}),
we have
\begin{eqnarray}
 &  & \big|\Phi_{z^{2},s}(\vec{R}_{i})\big>=V_{0}\big|\Omega_{0}^{s}(\vec{R}_{i})\big>,\label{omega1}\\
 &  & \big|\Phi_{xz,s}(\vec{R}_{i})\big>=\imath\frac{V_{1}}{\sqrt{2}}\big(\big|\Omega_{1}^{s}(\vec{R}_{i})\big>-\big|\Omega_{-1}^{s}(\vec{R}_{i})\big>\big),\label{omega2}\\
 &  & \big|\Phi_{yz,s}(\vec{R}_{i})\big>=\frac{V_{1}}{\sqrt{2}}\big(\big|\Omega_{1}^{s}(\vec{R}_{i})\big>+\big|\Omega_{-1}^{s}(\vec{R}_{i})\big>\big),\label{omega3}\\
 &  & \big|\Phi_{xy,s}(\vec{R}_{i})\big>=\imath\frac{V_{2}}{\sqrt{2}}\big(\big|\Omega_{2}^{s}(\vec{R}_{i})\big>-\big|\Omega_{-2}^{s}(\vec{R}_{i})\big>\big),\label{omega4}\\
 &  & \big|\Phi_{x^{2}-y^{2},s}(\vec{R}_{i})\big>=-\frac{V_{2}}{\sqrt{2}}\big(\big|\Omega_{2}^{s}(\vec{R}_{i})\big>+\big|\Omega_{-2}^{s}(\vec{R}_{i})\big>\big).\label{omega5}
\end{eqnarray}
We now move gears to the top-position case. Due to conservation of
angular momentum, $t_{l,s}^{(0,A)}$ and $t_{l,s}^{(0,B)}$ are non-zero
only for $l=z^{2}$, $t_{l,s}^{(0,A)}=V_{0}^{0A}=V_{pd\sigma}^{0A}$,
and $t_{l,s}^{(0,B)}=V_{0}^{0B}=V_{pd\sigma}^{0B}$. The explicit
expressions of $t_{l,s,j}^{(1,A)}$ are
\begin{eqnarray}
 &  & t_{z^{2},s,j}^{(1A)}=V_{0}^{1A},\label{t11A}\\
 &  & t_{xz,s,j}^{(1A)}=V_{1}^{1A}\frac{\imath}{\sqrt{2}}(e^{2\pi\imath j/3}-e^{-2\pi\imath j/3}),\label{t21A}\\
 &  & t_{yz,s,j}^{(1A)}=V_{1}^{1A}\frac{1}{\sqrt{2}}(e^{2\pi\imath j/3}+e^{-2\pi\imath j/3}),\label{t31A}\\
 &  & t_{xy,s,j}^{(1A)}=-V_{2}^{1A}\frac{\imath}{\sqrt{2}}(-e^{4\pi\imath j/3}+e^{-4\pi\imath j/3}),\label{t41A}\\
 &  & t_{x^{2}-y^{2},s,j}^{(1A)}=-V_{2}^{1A}\frac{1}{\sqrt{2}}(e^{4\pi\imath j/3}+e^{-4\pi\imath j/3}),\label{t51A}
\end{eqnarray}
The explicit expressions of $t_{l,s,j}^{(1,B)}$ are
\begin{eqnarray}
 &  & t_{z^{2},s,j}^{(1B)}=V_{0}^{1B},\label{t11B}\\
 &  & t_{xz,s,j}^{(1B)}=V_{1}^{1B}\frac{\imath}{\sqrt{2}}(-e^{-2\pi\imath j/3}+e^{2\pi\imath j/3}),\label{t21B}\\
 &  & t_{yz,s,j}^{(1B)}=-V_{1}^{1B}\frac{1}{\sqrt{2}}(e^{-2\pi\imath j/3}+e^{2\pi\imath j/3}),\label{t31B}\\
 &  & t_{xy,s,j}^{(1B)}=-V_{2}^{1B}\frac{\imath}{\sqrt{2}}(-e^{-4\pi\imath j/3}+e^{4\pi\imath j/3}),\label{t41B}\\
 &  & t_{x2-y^{2},s,j}^{(1B)}=-V_{2}^{1B}\frac{1}{\sqrt{2}}(e^{-4\pi\imath j/3}+e^{4\pi\imath j/3}).\label{t51B}
\end{eqnarray}
The constants of Eqs (\ref{t11A}-\ref{t51A}) are $V_{0}^{1A}=\sqrt{3}n_{1A}(1-n_{1A}^{2})V_{pd\pi}^{(1A)}-\frac{1}{2}n_{1A}(1-3n_{1A}^{2})V_{pd\sigma}^{(1A)}$,
$V_{1}^{1A}=\frac{1}{\sqrt{2}}\big(\sqrt{3}n_{1A}^{2}V_{pd\sigma}^{(1A)}+(1-2n_{1A}^{2})V_{pd\pi}^{(1A)}\big)\sqrt{1-n_{1A}^{2}}$,
and $V_{2}^{1A}=\frac{1}{\sqrt{2}}n_{1A}(1-n_{1A}^{2})\big(\frac{\sqrt{3}}{2}V_{pd\sigma}^{(1A)}-V_{pd\pi}^{(1A)}\big)$,
and by exchanging $A\rightarrow B$, for constants in (\ref{t11B}-\ref{t51B}).

$\Phi$-states of Eq (\ref{Tt}) can be write in terms of triangular
states \cite{Pachoud2014},

\begin{eqnarray}
 &  & \big|\Gamma_{m}^{s(1A)}(\vec{R}_{i})\big>=\sum_{j=0}^{2}e^{im2\pi j/3}\big|B,s,\vec{R}_{i}+\vec{a}_{j+1}\big>,\label{g1}\\
 &  & \big|\Gamma_{m}^{s(1B)}(\vec{R}_{i})\big>=\sum_{j=0}^{2}e^{im2\pi j/3}\big|A,s,\vec{R}_{i}+\vec{a}_{1}+\vec{\delta}_{j+1}\big>.\label{g2}
\end{eqnarray}
Here, $\vec{\delta}_{j}$ are given by $\vec{\delta}_{1}=-\vec{a}_{1}$,
$\vec{\delta}_{2}=-\vec{a}_{3}$, and $\vec{\delta}_{3}=-\vec{a}_{2}$.
States (\ref{g1}) and (\ref{g2}), similarly to states (\ref{hexagonal}),
have well defined angular momentum and satisfy $\big|\Gamma_{2}\big>=\big|\Gamma_{-1}\big>$,
and $\big|\Gamma_{-2}\big>=\big|\Gamma_{1}\big>$. In other words,
graphene does not support triangular states with $\big|m\big|=2$
\cite{Pachoud2014}. Finally, we find 
\begin{eqnarray}
 &  & \big|\Phi_{s,z^{2}}^{(1A)}(\vec{R}_{i})\big>=V_{0}^{1A}\big|\Gamma_{0}^{s(1A)}(\vec{R}_{i})\big>,\label{phi11A}\\
 &  & \big|\Phi_{s,xz}^{(1A)}(\vec{R}_{i})\big>=V_{1}^{1A}\frac{\imath}{\sqrt{2}}\big(\big|\Gamma_{1}^{s(1A)}(\vec{R}_{i})\big>-\big|\Gamma_{-1}^{s(1A)}(\vec{R}_{i})\big>\big),\nonumber \\
\label{phi21A}\\
 &  & \big|\Phi_{s,yz}^{(1A)}(\vec{R}_{i})\big>=V_{1}^{1A}\frac{1}{\sqrt{2}}\big(\big|\Gamma_{1}^{s(1A)}(\vec{R}_{i})\big>+\big|\Gamma_{-1}^{s(1A)}(\vec{R}_{i})\big>\big),\nonumber \\
\label{phi31A}\\
 &  & \big|\Phi_{s,xy}^{(1A)}(\vec{R}_{i})\big>=-V_{2}^{1A}\frac{\imath}{\sqrt{2}}\big(-\big|\Gamma_{-1}^{s(1A)}(\vec{R}_{i})\big>+\big|\Gamma_{1}^{s(1A)}(\vec{R}_{i})\big>\big),\nonumber \\
\label{phi41A}\\
 &  & \big|\Phi_{s,x^{2}-y^{2}}^{(1A)}(\vec{R}_{i})\big>=-V_{2}^{1A}\frac{1}{\sqrt{2}}\big(\big|\Gamma_{-1}^{s(1A)}(\vec{R}_{i})\big>+\big|\Gamma_{1}^{s(1A)}(\vec{R}_{i})\big>\big),\nonumber \\
\label{phi51A}
\end{eqnarray}
and, 
\begin{eqnarray}
 &  & \big|\Phi_{s,z^{2}}^{(1B)}(\vec{R}_{i})\big>=V_{0}^{1B}\big|\Gamma_{0}^{s(1B)}(\vec{R}_{i})\big>,\label{phi11B}\\
 &  & \big|\Phi_{s,xz}^{(1B)}(\vec{R}_{i})\big>=V_{1}^{1B}\frac{\imath}{\sqrt{2}}\big(-\big|\Gamma_{-1}^{s(1B)}(\vec{R}_{i})\big>+\big|\Gamma_{1}^{s(1B)}(\vec{R}_{i})\big>\big),\nonumber \\
\label{phi21B}\\
 &  & \big|\Phi_{s,yz}^{(1B)}(\vec{R}_{i})\big>=-V_{1}^{1B}\frac{1}{\sqrt{2}}\big(\big|\Gamma_{-1}^{s(1B)}(\vec{R}_{i})\big>+\big|\Gamma_{1}^{s(1B)}(\vec{R}_{i})\big>\big),\nonumber \\
\label{phi31B}\\
 &  & \big|\Phi_{s,xy}^{(1B)}(\vec{R}_{i})\big>=-V_{2}^{1B}\frac{\imath}{\sqrt{2}}\big(-\big|\Gamma_{1}^{s(1B)}(\vec{R}_{i})\big>+\big|\Gamma_{-1}^{s(1B)}(\vec{R}_{i})\big>\big),\nonumber \\
\label{phi41B}\\
 &  & \big|\Phi_{s,x^{2}-y^{2}}^{(1B)}(\vec{R}_{i})\big>=-V_{2}^{1B}\frac{1}{\sqrt{2}}\big(\big|\Gamma_{1}^{s(1B)}(\vec{R}_{i})\big>+\big|\Gamma_{-1}^{s(1B)}(\vec{R}_{i})\big>\big).\nonumber \\
\label{phi51B}
\end{eqnarray}


\begin{thebibliography}{10}
\bibitem{note_local}The relative strength of the proximity interactions at low energy is determined by the details of the hybridization $V$ between electronic states defined near the Dirac points ($\mathbf{k}=0$), which is independent of the electronic dispersion away from those points.
\bibitem{comment_energies}For convenience, the energy levels $\epsilon_{d_{l}}$ are defined
with respect to the on-site energy of carbon atoms.
\end{thebibliography}
\end{document}